\documentclass[%
 preprint,
 amsmath,amssymb,
 aps, physrev,
]{revtex4-2}

\usepackage{graphicx}
\usepackage{dcolumn}
\usepackage{bm}
\usepackage{comment} 

\begin{document}


\title{\textbf{Free Surface Enhancement of Droplet Rupture by Cavitation Bubble Collapse} 
}%

\author{Chenghao Xu}
\affiliation{Mechanical Science and Engineering, University of Illinois Urbana-Champaign, Illinois 61801, USA}
\author{Zhengyu Yang}%
\affiliation{Mechanical Science and Engineering, University of Illinois Urbana-Champaign, Illinois 61801, USA}
\author{Jie Feng}%
 \email{Contact author: jiefeng@illinois.edu}
\affiliation{Mechanical Science and Engineering, University of Illinois Urbana-Champaign, Illinois 61801, USA}
\affiliation{Materials Research Laboratory, University of Illinois Urbana-Champaign, Illinois 61801, USA}

\date{\today}

\begin{abstract}
The interaction between cavitation bubbles and surrounding droplets plays a central role in applications such as surface cleaning, ultrasonic emulsification, and therapeutic delivery. These processes depend on bubble-driven microjets that drive the deformation and breakup of the droplets, which are significantly influenced by geometric confinements. Here, we investigate the hydrodynamic interaction between cavitation bubbles and oil droplets within a thin water layer considering the coupling confinements of a free surface and a rigid wall. We reveal two distinct regimes of droplet response to cavitation bubble collapse: the rupture regime, where oil droplets fragment into smaller droplets, and the no-rupture regime, where the droplet remains intact. By deriving a non-dimensional Kelvin impulse to represent the momentum of the bubble-induced jet, we establish a scaling law that correlates the criterion for droplet rupture to a characteristic Weber number and the bubble-to-droplet size ratio for the first time. This framework delineates the rupture boundary and even extends to predict the rupture of particle-laden droplets driven by cavitation bubbles. Our findings reveal the hydrodynamic principles underlying the cavitation bubble-driven droplet rupture and provide predictive criteria for controlling performance in engineering and biomedical systems involving cavitation bubble dynamics.
\begin{description}
\item[Key words]
Cavitation dynamics, jetting, bubble-droplet interaction, droplet rupture
\end{description}
\end{abstract}

\maketitle


\newpage
\section{\label{sec:level1} Introduction}

Cavitation refers to the formation and collapse of vapor-filled bubbles in a liquid triggered by local pressure reductions or energy deposition~\cite{brennen_cavitation_nodate}. The occurrence of cavitation in multiphase complex liquids, including those laden with droplets, particles, or cells, underpins a wide spectrum of technological and biomedical processes, including ultrasonic cleaning~\cite{maisonhaute_surface_2002, chahine_modeling_2016, reuter_flow_2017}, emulsification in the food and pharmaceutical industries~\cite{califano_experimental_2014, mura_emulsion_2014, raman_cavitation_2022}, and drug delivery in biomedical engineering~\cite{kuznetsova_cavitation_2005, gac_sonoporation_2007, coussios_applications_2008, cattaneo_cyclic_2025, brennen_cavitation_2015}. The effectiveness of these applications is governed by the complex interactions between cavitation-induced flows and suspended particulates. Consequently, the bubble-particulate interactions have received great attention given the broad practical relevance, i.e., how high-speed microjets from collapsing bubbles, strong shear flows, and shock wave emissions deform, disrupt, or rupture nearby droplets, particles, or cells~\cite{ashoke_raman_microemulsification_2022, ren_interactions_2023, ren_particulate_2022, poulain_particle_2015, zhou_controlled_2012, yang_mechanisms_2020}.

In recent studies, the interaction between cavitation bubbles and droplets has attracted particular fundamental interest. For example, prior work has considered configurations of free-settling droplets~\cite{raman_cavitation_2022, ashoke_raman_microemulsification_2022}, droplets suspended on needles~\cite{stepisnik_perdih_revision_2019, orthaber_cavitation_2020, orthaber_cavitation_2020}, and droplets pendent from a solid surface~\cite{ren_interactions_2023}. These efforts have uncovered a wide spectrum of interfacial phenomena, including droplet deformation, bubble penetration, and emulsification. Yet, most of these studies were conducted in unbounded liquid domains or in the presence of a single nearby boundary (See Supplemental Material (SM), Sec. S1 Table~S1)
In contrast, realistic environments often impose multiple boundaries, such as free surfaces, solid walls, or thin liquid layers~\cite{zevnik_dynamics_2024, zwaan_controlled_2007, jalaal_laser-induced_2019}, that collectively shape bubble dynamics and droplet responses. 
This intricate interplay between cavitation bubbles and multiple boundaries not only adds to the complexity of cavitation dynamics but also expands its potential applications, such as mitigating cavitation erosion by trapping gas bubbles on top of the solid surface~\cite{gonzalez-avila_mitigating_2020, wei_manipulation_2024}, deforming single cells by generating tandem cavitation bubbles in microfluidics~\cite{yuan_cell_2015}.
Consequently, elucidating bubble-droplet interactions under multiple boundaries is essential to develop predictive models and to rationalize the design and control of cavitation-based technologies. 

Here, we conducted experiments on the interaction between a cavitation bubble and an oil droplet in a thin water layer, where a free surface and a rigid wall jointly influence bubble collapse. We showed that the presence of the free surface promotes droplet rupture, where droplet rupture can occur even at large bubble–droplet standoff distances from the bottom wall. By further combining the Kelvin impulse with the characteristic Weber number and bubble-to-droplet size ratio, we successfully developed a predictive scaling threshold that delineates rupture and no-rupture regimes observed in the experiments for both pure and particle-laden oil droplets.

\begin{figure}
    \centering
    \includegraphics[width=0.8\linewidth]{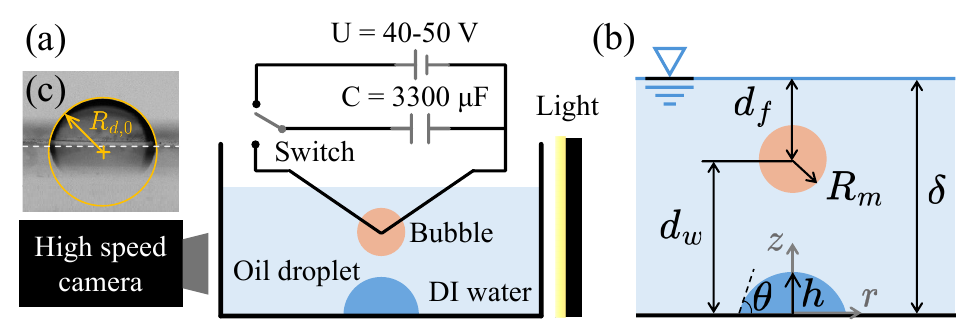}
    \caption{Experimental setup for the interaction between a cavitation bubble and an oil droplet in a thin liquid layer. (a) Schematic representation of the experimental configuration. (b) Zoomed-in schematic of the cavitation bubble and the droplet when the bubble expands to the maximum radius $R_m$, with definition of various length scales relevant to the experimental system. (c) Initial droplet shape showing that the droplet can be approximated by a spherical cap with an equivalent radius $R_{d,0}$.}
    \label{fig:ExperimentSetup}
\end{figure}

\section{\label{sec:level1} Experiments for droplet rupture by cavitation bubbles}

We performed controlled experiments as shown in Fig.~\ref{fig:ExperimentSetup}. Cavitation bubbles were generated by a spark created through the discharge of a capacitor, with the bubble size controlled by adjusting the capacitor's charging voltage (see Materials and Methods). The evolution of the cavitation bubble and its subsequent interaction with the droplet is recorded by high-speed imaging. Maintaining a fixed value of the equivalent droplet radius $R_{d,0}=2.67\pm0.13~\mathrm{mm}$ (see Fig.~\ref{fig:ExperimentSetup}), we observed that droplet rupture dynamics were controlled by varying three key geometric parameters: the distance from the center of the bubble to the bottom wall $d_w$, the thickness of the water layer $\delta$, and the maximum radius of the bubble $R_m$. 

\begin{figure}[htp]
    \centering
    \includegraphics[width=0.9\linewidth]{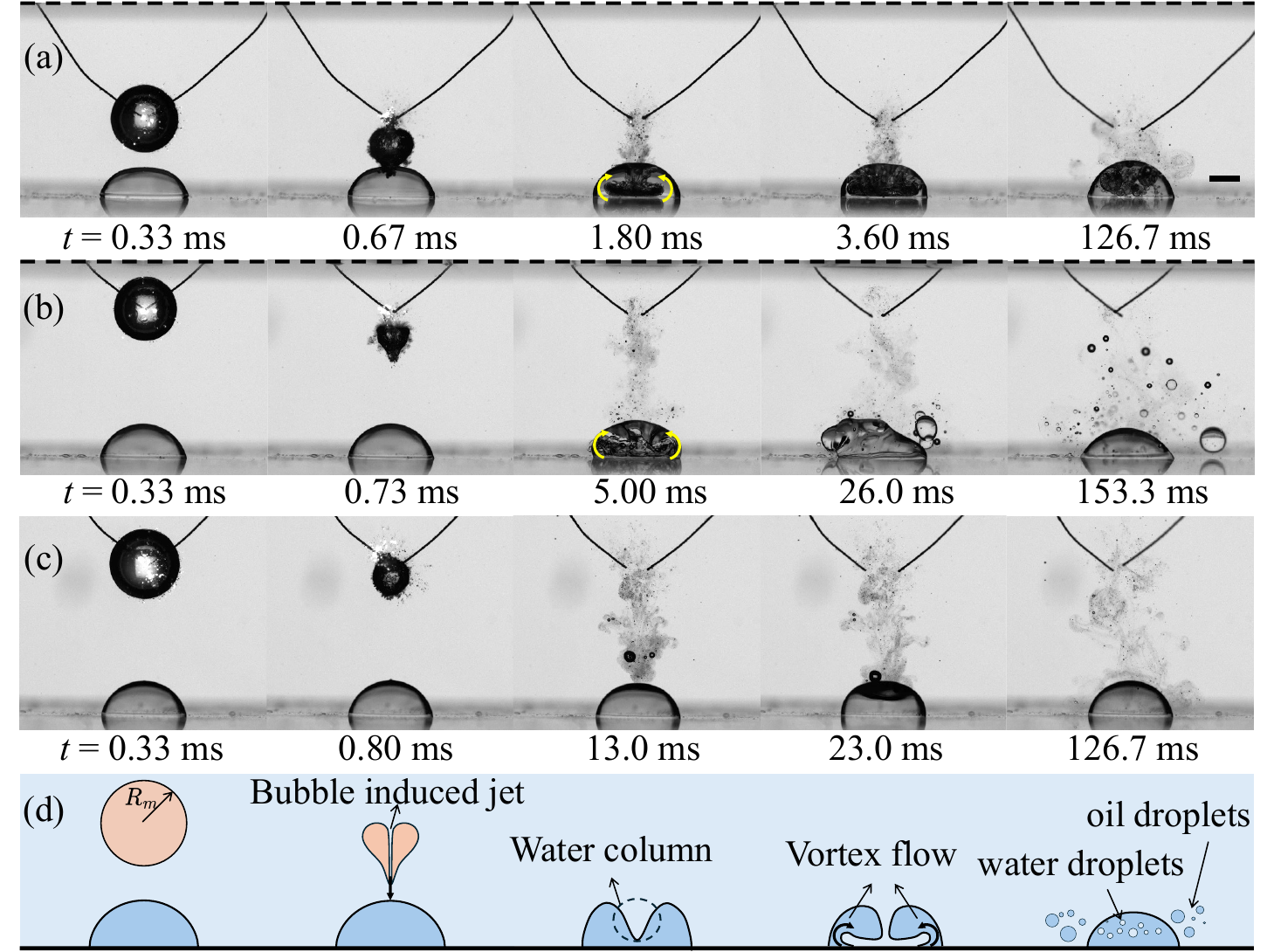}
    \caption{
    Interactions between a cavitation bubble and a silicone oil droplet under different conditions. (a) No-rupture regime at $\delta/R_m = 5.73$, $d_w/R_m = 2.36$, $R_m/R_{d,0} = 0.86$, with emulsified water trapped inside the droplet. (b) Rupture regime at $\delta/R_m = 6.00$, $d_w/R_m = 4.61$, $R_m/R_{d,0} = 0.85$. (c) High-speed images of cavitation bubble in a deep pool without free surface, with $d_w/R_m$ and $R_m/R_{d,0}$ close to those in (b). (d) Schematic illustrating how cavitation-induced flow interacts with a droplet, leading to droplet rupture. $t = 0$ is spark initiation; arrows show flows after the water column impacts the bottom wall at $t = 1.80\ \mathrm{ms}$ (a) and $t = 5.00\ \mathrm{ms}$ (b). The black dashed line in (a, b) marks the free surface. Scale bar: 2 mm. 
    The black dash line in (a) and (b) indicates the free surface. The scale bar represents 2 mm. 
    }
    \label{fig:ExperimentResults}
\end{figure}

Two distinct behaviors of oil droplets are identified in response to the cavitation bubble: the rupture and the no-rupture regimes, as illustrated in Fig.~\ref{fig:ExperimentResults}(a) and (b), respectively.
Fig.~\ref{fig:ExperimentResults}(a) shows the motion of a cavitation bubble that induces no droplet rupture, under conditions $\delta/R_m=5.73$, $d_w/R_m=2.36$, $R_m/R_{d,0}=0.86$ (see Movie S1).  After the initial collapse and the subsequent rebound, the cavitation bubble generates a jet, initiating the downward motion due to the initial impulse, which penetrates into the oil droplet and forms an inward water column. This water column travels downward through the droplet and approaches the rigid wall at $t=1.4~\mathrm{ms}$. Upon reaching the wall, the water column begins to extend radially outward, creating a vortex flow, as indicated by the yellow arrows at $t=1.80~\mathrm{ms}$. During this process, the water column pinches off from the surrounding water layer at the bottom of the oil droplet. As the water column continues to spread radially, it stretches and divides into smaller water droplets. However, because of the relatively weak momentum of the bubble-induced flow, the resulting hydrodynamic force is insufficient to overcome the capillary force at the oil-water interface. Thus the oil droplet remains intact with several smaller water droplets trapped inside, as shown at $t=126.7~\mathrm{ms}$. 

In comparison, Fig.~\ref{fig:ExperimentResults}(b) shows the rupture of a silicone oil droplet caused by the interaction with bubble-induced flow, under the conditions $\delta/R_m = 5.53$, $d_w/R_m = 2.49$, and $R_m/R_{d,0} = 0.85$ (see Movie S2). The cavitation bubble expands, collapses, and moves toward the bottom wall, similar to that in Fig.~\ref{fig:ExperimentResults}(a) until the bubble-induced flow approaches the oil drop. The bubble-induced jet penetrates the droplet, drives a downward-moving water column, and circulates internally, as indicated at $t=5.00~\mathrm{ms}$. This internal circulation generates strong stresses within the droplet, destabilizing the oil-water interface and inducing substantial deformation. The combined hydrodynamic effects eventually lead to the rupture of the droplet. With multiple smaller oil droplets detached from the parent droplet, a noticeable reduction in mass can be observed. This process highlights the role of bubble-induced flow dynamics in destabilizing the oil-water interface and driving droplet fragmentation under these specific conditions, as illustrated schematically in Fig. \ref{fig:ExperimentResults}(d). To highlight the effect of the free surface, we further conducted baseline experiments under the same conditions but without the presence of a free surface, as shown in Fig.~\ref{fig:ExperimentResults}(c) (see Movie S3). In this case, the collapse of the bubble generates a weaker jet, indicating that the downward momentum is insufficient to rupture the droplet. The bubble-induced flow only causes a mild deformation of the oil droplet without the free surface. This comparison pinpoints the critical role of the free surface in enhancing the downward momentum of the bubble-induced flow, which contributes to the rupture and fragmentation of the oil droplet.

Building on the above comparative observations, the rupture of the near-wall droplet arises from the inertial effect of a bubble jet which induces a water flow directed toward the wall, overcoming the stabilizing surface tension effect and causes droplet disintegration.
The key factor determining whether the droplet will rupture is the momentum of the cavitation bubble after it collapses.
Comparing Fig.~\ref{fig:ExperimentResults}(b) and (c), we observe that the bubble-induced jet travels much more slowly when the free surface is absent. The free surface promotes a bubble-induced jet with higher momentum towards the wall, resulting in higher kinetic energy to impact the droplet and cause its rupture. 
To quantify the inertial effect of the bubble jet, we will next model the momentum of the cavitation bubble and evaluate its impact on the bubble-droplet interaction.

\section{\label{sec:level1} Modeling flow field and Kelvin impulse}
The experimental observations demonstrate that the multiple-boundary confinement critically influences the bubble-induced flow, and consequently, the bubble-droplet interaction. Specifically, the initial distance between the cavitation bubble and the boundaries decides how much the momentum of the bubble carries~\cite{blake_cavitation_nodate}, which further determines key phenomena in the bubble-droplet, such as the pinch-off of small oil droplets from the original droplet (rupture regime) or the entrapment of water droplets within the oil droplet (no-rupture regime)
~\cite{ren_interactions_2023}. To quantitatively describe the associated flow and capture the dynamics of the cavitation bubble near multiple boundaries, we adopt a theoretical framework that combines the method of images for potential flow \cite{noauthor_underwater_nodate, best_estimate_1994} and the Kelvin impulse approach \cite{blake_note_1982, blake_cavitation_nodate, supponen_scaling_2016}, which has proved a solid method to characterize the momentum of cavitation bubbles in previous studies of their asymmetric collapse near a boundary \cite{supponen_scaling_2016, ren_interactions_2023}.

\begin{figure}
    \centering
    \includegraphics[width=0.6\linewidth]{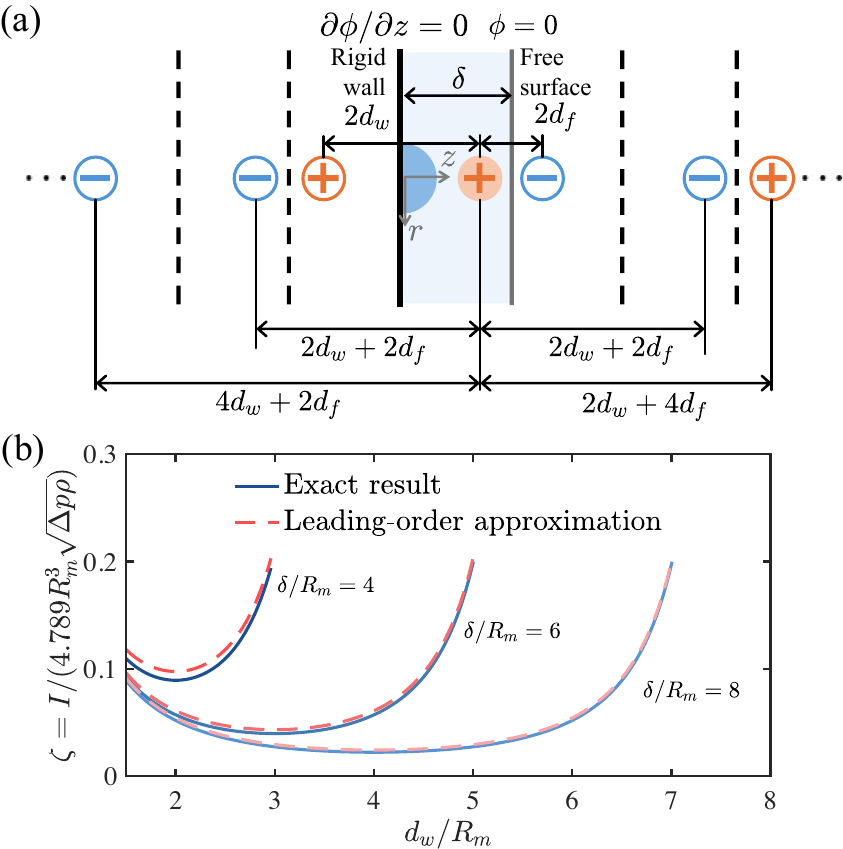}
    \caption{(a) Physical model based on the the method of image. To satisfy the boundary conditions, two infinite sets of image bubbles are employed to represent the contribution to the velocity potential. The orange circle with a plus sign represents the original bubble as the source of $m(t)$. Uncolored circles represent the two infinite sets of image bubbles, where circles with plus signs denote image sources, and circles with minus signs denote image sinks. (b) Comparison between the leading-order approximation of $\zeta$, which considers only the closest source and the closest sink, and the exact solution. Results are shown for three different ratios: $\delta/R_m = 4$ (minimum), $\delta/R_m = 6$ (median), and $\delta/R_m = 8$ (maximum) used in the experiments.}
    \label{fig:ImageMethod}
\end{figure}

\subsection{\label{sec:level2} Flow field based on the method of images}
We use the method of images commonly adopted for modeling of cavitation bubble motion during the initial expanding-collapsing process of the bubble \cite{li_vertically_2023, tagawa_bubble_2018}. The method of images depends on the potential flow assumptions, requiring the flow to be incompressible, inviscid and irrotational. Given the short characteristic timescale associated with the evolution of the cavitation bubble, both the oil droplet and the surrounding water layer can be approximated as incompressible fluids~\cite{han_interaction_2022, ren_interactions_2023}. The Reynolds number of the cavitation bubble, defined as $Re_b = R_m \sqrt{\Delta p / \rho_w} / \nu_w$, ranges from $1.7 \times 10^4$ to $3.5 \times 10^4$, indicating that viscous effects are negligible in our case. Here, $\Delta p = p_\infty - p_v$ represents the pressure difference driving the bubble collapse, where $p_\infty = 1.01\times10^5 ~\mathrm{Pa}$ is the far-field pressure and $p_v = 2.3 \times 10^3 ~\mathrm{Pa}$ is the vapor pressure of the water. The density and kinematic viscosity of water are taken as $\rho_w = 1 \times 10^3~\mathrm{kg\,m}^{-3}$ and $\nu_w = 1 \times 10^{-6}~\mathrm{m^2\,s^{-1}}$, respectively. Under the assumptions of incompressible, inviscid and irrotational flow, we can thus use the potential flow theory to determine the velocity potential $\phi$~\cite{blake_cavitation_nodate}, which allows computation of the Kelvin impulse to describe bubble behaviors influenced by multiple boundaries.

This approach is schematically illustrated in Fig.~\ref{fig:ImageMethod}(a). The cavitation bubble is represented as a point source, illustrated by a solid orange circle, with a source strength $m(t)$ varying in time $t$. The velocity potential satisfies the Laplace equation $\nabla^2\phi=0$ \cite{blake_kelvin_1988} and the boundary conditions, namely zero velocity in the $z$ direction on the rigid bottom wall expressed as $\partial \phi/\partial z=0$ and zero potential on the free surface expressed as $\phi=0$. To fulfill the boundary conditions, two infinite series of image bubbles organized in a defined order are introduced \cite{huang_wall_2024}, as demonstrated in Fig.~\ref{fig:ImageMethod}(a). These image bubbles alternate as two sources and two sinks with respective strengths $m(t)$ and $-m(t)$, illustrated schematically as hollow orange circles labeled with plus signs for image sources and hollow blue circles labeled with minus signs for image sinks. We also neglect the effect of the droplet-water boundary condition due to the similar densities of water ($\rho_w=1\times10^3~\mathrm{kg}~\mathrm{m}^{-3}$) and the droplet (silicone oil AR20, $\rho_o = 1.00\mathrm{-}1.02\times10^3~\mathrm{kg}~\mathrm{m}^{-3}$), as the ratio of the momentum contribution between the water-oil interface and the other interfaces has been characterized in the literature to scale with the Atwood number $At = (\rho_w - \rho_o)/(\rho_w + \rho_o)$ (see SI Appendix Section II for a detailed discussion).

Using the method of images, we obtain the flow field in the water layer induced by the bubble (see SI Appendix Section II). The induced velocity at the bubble, directed in the $z$-direction due to axisymmetry, is expressed as $\nabla\phi(a) = - m(t)\Phi(a)/64\pi\delta^2\mathbf{e}_z$, where the function $\Phi(a)$ is defined as
\begin{equation}
    \Phi(a) = \varphi\left(\frac{a}{2}\right) 
            + \varphi\left(\frac{1}{2} - \frac{a}{2}\right)
            - \varphi\left(\frac{1}{2} + \frac{a}{2}\right) 
            - \varphi\left(1 - \frac{a}{2}\right).
\end{equation}
Here, $\varphi(z) = \Sigma_{n=0}^\infty1/(z+n)^2$ denotes the trigamma function \cite{abramowitz1965handbook}, $\delta$ is the thickness of the water layer, and the dimensionless parameter $a$ is given by $a = d_w/\delta$, where $d_w$ represents the distance from the bubble center to the rigid wall.   

\subsection{\label{sec:level2} Calculation of non-dimensional Kelvin impulse}
With the flow field described by the method of images, we can calculate the Kelvin impulse to describe the apparent inertia of the bubble \cite{blake_kelvin_1988}, which can be derived as~\cite{best_estimate_1994, blake_cavitation_2015, supponen_scaling_2016}
\begin{equation}
    I = \rho_w\int_{0}^{2T_c}-m(t)\nabla\phi_z dt = \frac{\rho_w\Phi(a)}{64\pi\delta^2}\int_{0}^{2T_c}m^2(t)dt,
\end{equation}
where $T_c$ is the bubble collapse time given by $T_c=\xi R_m(\rho_w/\Delta p)^{1/2}$, where $\xi\approx 0.914$ is the Rayleigh factor~\cite{supponen_scaling_2016}. 
We further derive a non-dimensional form of the Kelvin impulse as (see SI Appendix Section III for details)
\begin{equation}
    \zeta = \frac{I}{4.789\, R_m^3\sqrt{\Delta p\,\rho_w}} = 0.049\left(\frac{R_m}{\delta}\right)^2\Phi\left(\frac{d_w}{R_m}\cdot\frac{R_m}{\delta}\right).
\end{equation}

Fig.~\ref{fig:ImageMethod}(b) presents the dimensionless Kelvin impulse $\zeta$ of a cavitation bubble generated within the water layer as a function of the dimensionless parameter $d_w/R_m$. We introduce $\zeta_1$, the leading-order approximation of the dimensionless Kelvin impulse, which considers only the contributions from one image sink located directly above the free surface and one image source directly below the rigid wall as
\begin{equation}
    \zeta_1 = 0.195\left(\frac{R_m^2}{d_w^2} + \frac{R_m^2}{(\delta-d_w)^2}\right).
\end{equation}
The maximum deviation between this leading-order approximation and the exact solution is only $8.4\%$, regardless of the value of $R_m / \delta$. It shows that the contribution of sources and sinks far away from the original bubble decays over space and can be safely neglected. Due to its computational simplicity and acceptable accuracy, the leading-order approximation $\zeta \approx \zeta_1$ will be used for all subsequent analyses and calculations.

\begin{figure*}
    \centering
    \includegraphics[width=0.9\linewidth]{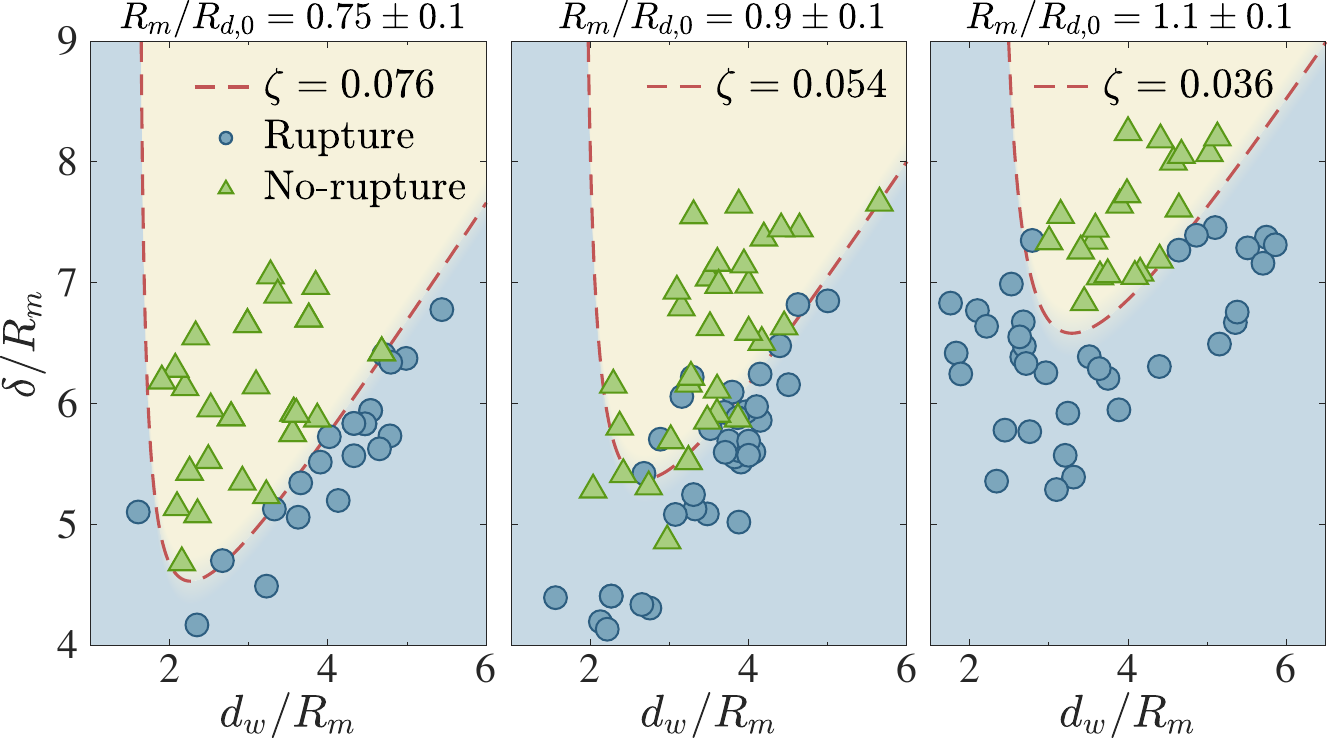}
    \caption{Regime maps of droplet rupture in the parameter space of $d_w/R_m$ and $\delta/R_m$ for different bubble-to-droplet radius ratios (a) $R_m/R_{d,0} = 0.75 \pm 0.1$, (b) $R_m/R_{d,0} = 0.9 \pm 0.1$, and (c) $R_m/R_{d,0} = 1.1 \pm 0.1$. The red dashed lines indicate the critical non-dimensional Kelvin impulses obtained by fitting, which define the threshold delineating the rupture and non-rupture regimes of the oil droplets.
    }
    \label{fig:PhaseDiagram}
\end{figure*}

\section{\label{sec:level1} Predicting droplet rupture with Kelvin impulse}

Fig.~\ref{fig:PhaseDiagram} presents the regime maps of the bubble-droplet interactions in the parameter space defined by dimensionless wall distances $d_w/R_m$ and dimensionless water layer thicknesses $\delta/R_m$ at different bubble-to-droplet size ratios $R_m/R_{d,0}$. In Fig.~\ref{fig:PhaseDiagram}(a), at a fixed $R_m/R_{d,0} = 0.75 \pm 0.1$, the rupture regime and the no-rupture regime are distinct on our regime map. We observe that a reduction of the dimensionless water layer thickness $\delta/R_m$ promotes the rupture of droplet, emphasizing the role of a free surface in propelling the cavitation bubble. Specifically, as $d_w$ becomes closer to $\delta$, droplet rupture becomes more likely even at a high $\delta/R_m$, consistent with Fig.~\ref{fig:ExperimentResults}(a-b) which shows that the droplet ruptures when the bubble position approaches the free surface. 

Additionally, we show that a constant non-dimensional Kelvin impulse $\zeta$, denoted by the dashed curve, serves as an effective criterion between the rupture and no-rupture regimes. The rupture of the droplet predominantly occurs when the non-dimensional Kelvin impulse is larger than the critical value, where the bubble-induced jet exhibits a higher momentum due to the stronger influence from the multiple boundaries, especially the free surface. Furthermore, Fig.~\ref{fig:PhaseDiagram}(b-c) illustrates that the critical $\zeta$ required for droplet rupture decreases as the ratio between the maximum bubble radius and the initial equivalent droplet radius $R_m/R_{d,0}$ increases. The trend can be attributed to the increase in kinetic energy of the bubble-induced jet with the bubble radius. As the bubble size increases, the collapsing cavitation produces a jet of larger volume, which generates a stronger bubble-induced flow with greater kinetic energy at the same $\zeta$. Consequently, the critical energy to rupture the droplet can be achieved at a smaller $\zeta$. Our findings demonstrate that the non-dimensional Kelvin impulse can be used as a critical parameter that determines whether a cavitation bubble can rupture the droplet. 

\begin{figure}[h]
    \centering
    \includegraphics[width=0.9\linewidth]{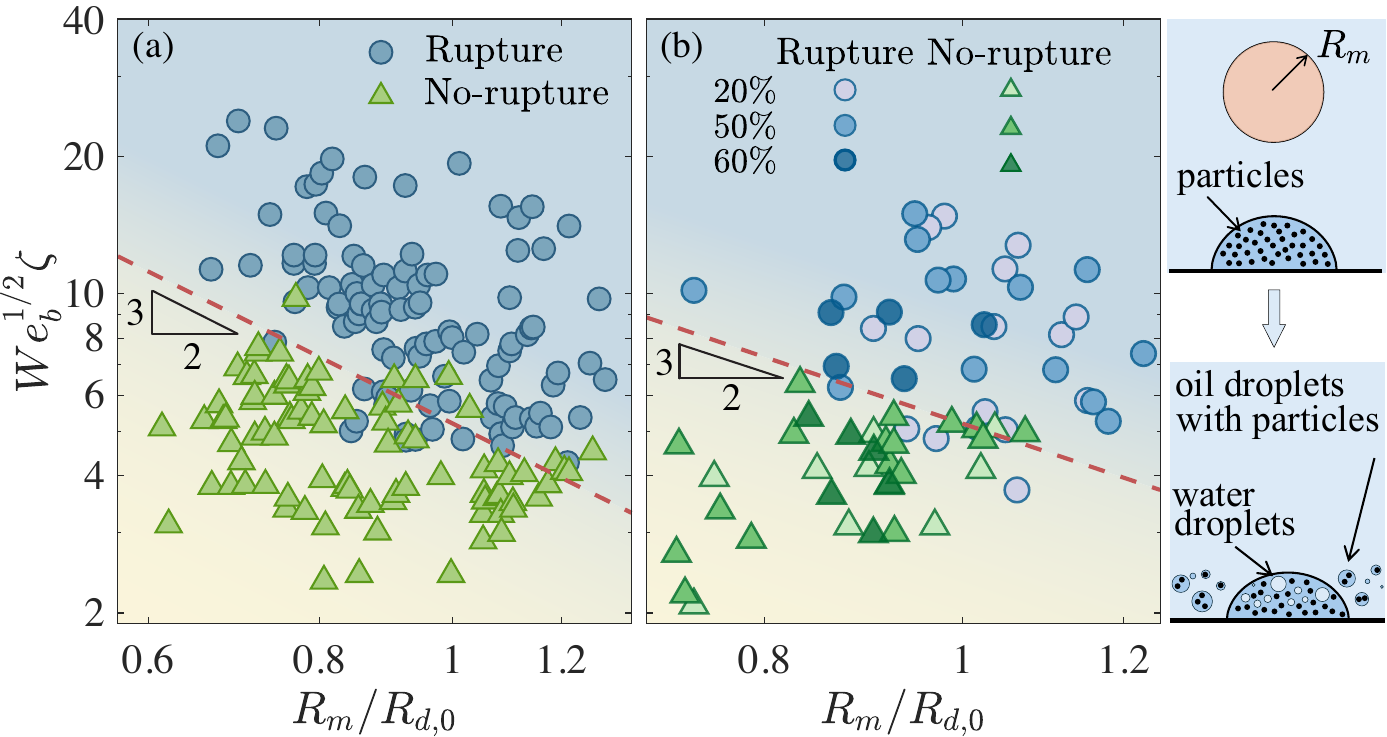}
    \caption{
    Regime maps of droplet rupture between a cavitation bubble and (a) an oil droplet and (b) a particle-laden oil droplet at volume fractions of 20\%, 50\%, and 60\% in a water layer. The dashed line represents the rupture threshold for both oil droplets, described by the relationship $We_b^{1/2}\zeta = 5.2\left(R_m/R_{d,0}\right)^{-3/2}$. The right panels in (b) show schematics of the rupture process of particle-laden oil droplets caused by cavitation bubbles.
    }
    \label{fig:CriticalKelvinImpulse}
\end{figure}

Next, we perform a scaling analysis based on energy balance to predict the droplet rupture. We review the process of droplet deformation and rupture induced by cavitation bubble shown in Fig.~\ref{fig:ExperimentResults}(d). Initially, bubble migration induces a downward flow of water, which penetrates the oil droplet and forms a thick inward water column.The high-speed inward water column impacts the bottom surface and expands radially outward, introducing strong stretching and shearing that significantly deform the droplet, and ultimately resulting in the droplet breakup. The droplet breakup is attributed to large deformation of the droplet, which occurs when sufficient kinetic energy of the bubble-induced jet is transferred into the excess surface energy. Therefore, to predict the critical conditions where droplet rupture occurs, we consider an energy argument that the kinetic energy of the inward water column should overcome the required surface energy to rupture the droplet. The kinetic energy of the water column $E_k$ is estimated as $\frac{1}{2}M_w u_w^2$ \cite{ren_interactions_2023}, where $M_w$ is the mass of the water column and $u_w$ is the characteristic velocity of the water column. The mass of the water column is approximated as $M_w \approx \rho_w \pi R_m^2 h_w$, where the height of the water column can be estimated as $h_w \approx R_{d,0}$. It is reasonable to approximate the linear momentum of the water column by the Kelvin impulse of the cavitation bubble, which gives a characteristic water jet velocity of $u_w \approx I/M_w$. Therefore, the kinetic energy becomes 
\begin{equation}
    E_k \approx \frac{1}{2}M_w u_w^2 \approx \frac{\zeta^2\left(4.789R_m^3\rho_w u_0\right)^2}{2\rho_w \pi R_m^2 R_{d,0}},
\end{equation}
where $u_0=\sqrt{\Delta p/\rho_w}$. We estimate the required surface energy $E_{s,d} \propto \gamma_{ow} R_{d,0}^2$~\cite{quetzeri2021impact, ren_interactions_2023}. Here, $\gamma_{ow}$ is the interfacial tension at the oil-water interface. With the criterion that the droplet could rupture only if $E_k \gtrsim E_{s,d}$, we obtain the scaling law predicting the critical non-dimensional Kelvin impulse as:
\begin{equation}
    \zeta_c \propto \left(\frac{R_m}{R_{d,0}}\right)^{-\frac{3}{2}}\left(\frac{\rho_w u_0^2 R_m}{\gamma_{ow}}\right)^{-\frac{1}{2}}.
\end{equation}
We define bubble Weber number as $We_b = \rho_w u_0^2 R_m/\gamma_{ow}$, and the above criterion is written as
\begin{equation} \label{eq:Webzeta}
    We_b^{\frac{1}{2}}\zeta_c \propto \left(\frac{R_m}{R_{d,0}}\right)^{-\frac{3}{2}}.
\end{equation}

This scaling provides a quantitative criterion for determining the droplet rupture threshold as a function of the bubble-to-droplet size ratio, as well as the Weber number of the bubble. We plot the regime map of droplet rupture in the parameter space of $We_b^{1/2} \zeta$ and $R_m/R_{d,0}$. The dashed line, represented by Eq.~(\ref{eq:Webzeta}), well delineates the rupture and no-rupture regimes, confirming the accuracy of our proposed scaling law. Specifically, the scaling law highlights that the droplet rupture is governed by the interplay between the momentum of the bubble, which drives deformation, and the interfacial tension between water and the droplet, which resists rupture. Notably, the proximity of the bubble to the free surface enhances the non-dimensional Kelvin impulse $\zeta$, resulting in a stronger momentum that facilitates drop rupture. Furthermore, the bubble-to-droplet size ratio $R_m/R_{d,0}$ acts as a geometric factor influencing the energy balance, with smaller droplets more easily ruptured when the non-dimensional Kelvin impulse and the bubble Weber number are held constant.


In realistic fluid environment, cavitation often occurs in fluid media containing solid particles. Cavitation-induced rupture of particle-laden droplets is of substantial interest encompassing applications such as slurry fuel atomization~\cite{gvozdyakov_improvement_2021, noh_atomization_2022}, spraying of functional coatings~\cite{suhag_film_2020}, and oil processing~\cite{stebeleva_application_2021}. Therefore, we further investigate the rupture of oil droplets containing spherical microparticles in volume fractions of $20\%$, $50\%$, and $60\%$ as shown in Fig.~\ref{fig:CriticalKelvinImpulse}(b), to evaluate the feasibility of cavitation-bubble-driven atomization of particle-laden fluids. Our results show that the proposed scaling law (Eq.~\ref{eq:Webzeta}) robustly distinguishes between rupture and no-rupture regimes even in the presence of microparticle-laden droplets.

\section{Conclusion}
In this study, we investigated the interaction between cavitation bubbles and oil droplets within a thin water layer, taking into account the effects of a free surface and a rigid wall. We identified two distinct regimes of droplet behavior: the rupture regime, where the oil droplet fragments into smaller daughter droplets, and the no-rupture regime, where the water flow driven by the bubble collapse enters the oil droplet but cannot breakup the oil-water interface of the droplet. These regimes are governed by the momentum of the cavitation bubble quantified by non-dimensional Kelvin impulse, which is determined by two non-dimensional geometric parameters: the distance from the bubble to the wall $d_w/R_m$ and the thickness of the water layer $\delta/R_m$. Our results revealed that the key mechanism driving droplet rupture is energy transfer from the bubble-induced jet to the droplet. When the momentum of the cavitation bubble exceeds a critical threshold, the induced hydrodynamic stresses are sufficient to overcome the surface tension of the droplet, resulting in rupture. This threshold was shown to decrease as the bubble-to-droplet size ratio increases. We also predict the parametric threshold to distinct regimes of rupture and no-rupture dynamics, using the Weber number, the Kelvin impulse, and the bubble-to-droplet size ratio. Additionally, we find that the scaling can be extended to particle-laden droplets. This observation indicates that our results capture the fundamental mechanism underlying bubble-droplet interaction dynamics, further highlighting the generality and robustness of the proposed scaling law.

Our results serve as a first step toward a predictive framework for bubble–droplet interactions in multiply confined geometries. With demonstrated robustness across a broad parameter range, our results offer potential engineering guidelines in applications like ultrasonic cleaning. Future investigations could extend the scope to cavitation in non-Newtonian media, which are ubiquitous in industrial and biomedical configurations. Incorporating different rheological properties such as shear-thinning, shear-thickening, and viscoelasticity will be essential to establish realistic rupture thresholds. In closing, we believe that our findings of bubble-droplet interaction under multiple boundaries provide new insights into cavitation-driven deformation and rupture, with broad relevance to techniques involving bubble-induced fragmentation in industry and biomedical applications.

\begin{acknowledgments}
C.X., Z.Y. and J.F. acknowledge partial support by the NSF under grant No. CBET 2426809 and the research support award RB24105 from the University of Illinois Urbana-Champaign. We also thank Zibo Ren for the helpful discussions.
\dots.
\end{acknowledgments}


\bibliography{Reference}

\end{document}